\documentclass[journal]{IEEEtran}

\usepackage{amssymb}
\usepackage{amsmath,epsfig,amssymb,verbatim}
\usepackage{cite}

\begin{document}

\title{Terahertz Communications for Massive Connectivity and Security in 6G and Beyond Era}

\author{Nan Yang and Akram Shafie\vspace{-5mm}

\thanks{The authors are with the Australian National University, Australia.}}

\maketitle

\begin{abstract}
Terahertz (THz) communications (THzCom) has experienced a meteoric rise of interest, due to its benefits for ultra-high data rate transmission in the sixth generation (6G) and beyond era. Despite so, the research on exploring the potential of THzCom for other performance targets anticipated by 6G, including massive connectivity and security, is still in its infancy. In this article, we start with briefly describing the unique peculiarities of THz channels, and then discuss theoretical frameworks to facilitate the analysis and design of THz transmission for achieving massive connectivity and security. Then we discuss promising spectrum management strategies, including the exploration of multiple THz transmission windows and frequency reuse with multiplexing and signal processing, to substantially increase the number of supported users and identify to-be-tackled challenges. We further present important research directions based on the principles of physical layer security, such as new spectrum allocation policies and beamforming algorithms, to fight against eavesdropping in THzCom systems, ushering in secure THzCom systems.
\end{abstract}

\section{Introduction}

Our society is in the era of mobile wireless Internet with explosive big data. While the fifth generation (5G) wireless systems are being deployed worldwide, the telecommunications community has identified emerging use cases and technologies needed for an intelligent information society in the 2030s \cite{Tataria_ProcIEEE_2021}, most of which have quality-of-service (QoS) requirements far beyond what 5G can offer. For example, as a promising technology allowing doctors to perform remote surgeries, holographic telepresence often requires vast amounts of data (e.g., holographic images from multiple viewpoints) to be instantly communicated at data rates of several hundred gigabits per second or beyond one terabit per second (Tbps) \cite{Tataria_ProcIEEE_2021}, which is 10--100 times higher than the maximum data rate of 5G. To satisfy such unprecedented QoS requirements, the global research community and industry has reached a consensus to exploit frequencies at the terahertz (THz) band (0.1--10 THz) for the sixth generation (6G) and beyond wireless systems \cite{Akyildiz_TCOM_2022}. Particularly, the relatively low THz band (e.g., 0.1--0.3 THz) is envisioned to be regulated for use in 6G \cite{Petrov_CM_2020}, while the higher THz band (e.g., 0.3--1 THz) is anticipated to be adopted in beyond 6G.

\begin{figure*}[!t]
\centering
\includegraphics[width=1.85\columnwidth]{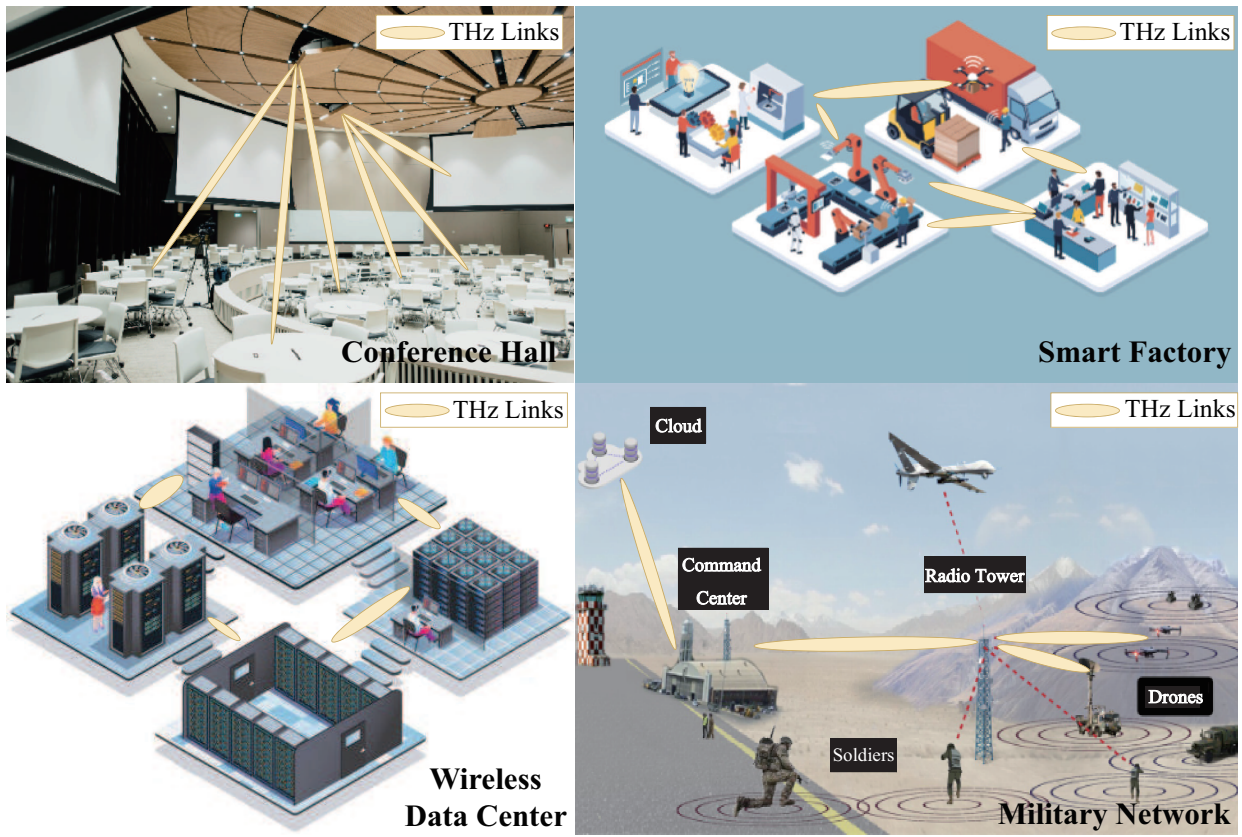}
\vspace{-2mm}
\caption{Massive connectivity and security oriented THzCom in 6G and beyond wireless applications.}\label{Fig:THzComApplications}
\vspace{-2mm}
\end{figure*}

The THz band has enormous potential to meet the demands of 6G, such as its far richer spectrum resources (e.g., tens up to hundreds of gigahertz (GHz) bandwidth) than the millimeter wave (mmWave) band adopted by 5G, and its capability to accommodate thousands of miniature antennas \cite{Akyildiz_TCOM_2022}. Thanks to the recent advances in THz hardware and the opening of the THz spectrum for research \cite{Petrov_CM_2020}, designing THz communication (THzCom) strategies for 6G and beyond wireless systems has become an emerging trend. From a communication and signal processing perspective, rapidly increasing efforts have been devoted to THzCom for achieving the QoS target of ultra-high data rate, aiming to realize promising 6G applications, such as Tbps WiFi for conferences and smart offices and Tbps links in wireless data centers, as shown in Fig. \ref{Fig:THzComApplications}.

It is worth noting that the ultra-high data rate is not the sole QoS requirement of 6G that THzCom can potentially satisfy. Particularly, the abundant spectrum resources at the THz band and the frequency- and distance-dependent nature of THz spectrum have substantial potential to meet other QoS requirements, including \emph{massive connectivity} and high \emph{security}. For example, THzCom-aided 6G systems can be adopted to serve a huge number of users with reasonably high data rate requirements, e.g., indoor and outdoor mega-sporting events (e.g., Olympic Games) and smart factories with massive wirelessly connected machine-type devices, as shown in Fig. \ref{Fig:THzComApplications}.
Also, THzCom-aided 6G systems must guarantee and maintain a very high level of secrecy, if used in wireless data centers and military communication networks, displayed in Fig. \ref{Fig:THzComApplications}. Against this background, our article discuss the advancements, opportunities, and challenges of using THzCom to support massive connectivity and security in the 6G and beyond wireless systems, with the primary focus on spectrum management and signal processing. Such discussion makes our article substantially different from existing articles focusing on ultra-fast THz transmission.

In this article, we first briefly describe the characteristics of THz channels, based on which the theoretical frameworks to analyze the connectivity and security performance of THzCom systems are investigated. Then, we identify new ways to achieve massive connectivity by exploring the rich THz spectrum resources and devising multiplexing and signal processing schemes. Furthermore, based on the principles of physical layer security, we delve into novel spectrum allocation policies and signal processing algorithms to safeguard THzCom systems. For both massive connectivity and security, we discuss open problems and future research directions to facilitate the establishment of theoretical frameworks and design of spectrum management and signal processing strategies.

\begin{figure*}[t]
\centering
\includegraphics[width=1.95\columnwidth]{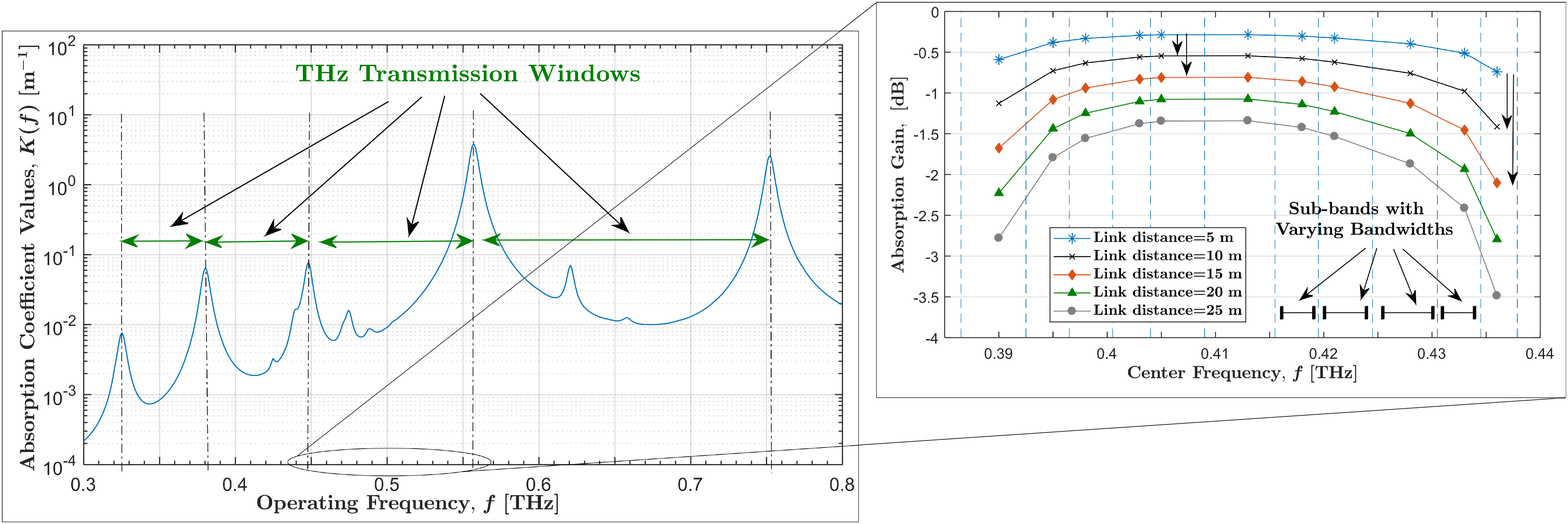}
\vspace{-4mm}
\caption{THz transmission windows with frequency- and distance-dependent molecular absorption loss.}\label{Fig:THzBand}\vspace{-2mm}
\end{figure*}

\section{THz Channel Characteristics and Performance Evaluation Frameworks}

One fundamental challenge to fully unleash the potential of THzCom is to understand the unique characteristics of THz channels that are not encountered at lower frequencies and then apply such characteristics into the analysis, design, and development of THzCom systems. Comparing to the channels accommodating the current wireless systems, the signals transmitted over THz channels experience very high spreading loss, frequency- and distance-dependent molecular absorption loss, exhibit high sparsity, and are very vulnerable to blockages, described as follows:

\textbf{Spreading Loss:} In wireless propagation, the spreading loss increases quadratically with frequency when the apertures of antennas, with frequency-independent gains, decrease at both ends of a wireless link. Given very high THz frequencies, the signal propagated over a THz link is impacted by severe spreading loss, e.g., the free space spreading loss experienced by a THz signal transmitted over ten meters is over $100~\textrm{dB}$ \cite{Guan_TTST_2019}.

\textbf{Molecular Absorption Loss:} The THz band is uniquely characterized by the molecular absorption loss, resulted from the phenomenon that THz signal energies are absorbed by water vapour and oxygen molecules in the propagation medium. The severity of molecular absorption loss depends on the operating frequency and transmission distance. Particularly, some THz frequencies are very sensitive to molecular absorption. Thus, molecular absorption peaks exist at such frequencies, where the THz signal energy profoundly decreases. This existence creates distance-dependent transmission windows (TWs), as illustrated in Fig. \ref{Fig:THzBand}. Within TWs, drastically varying molecular absorption loss is experienced and this variation increases with the transmission distance, as shown in Fig. \ref{Fig:THzBand}. The molecular absorption loss is a must-consider property for analyzing and designing THzCom systems, where the advantages of multiple TWs and their frequency- and distance-dependencies need to be explored.

\begin{figure*}[!t]
\centering
\includegraphics[width=2\columnwidth]{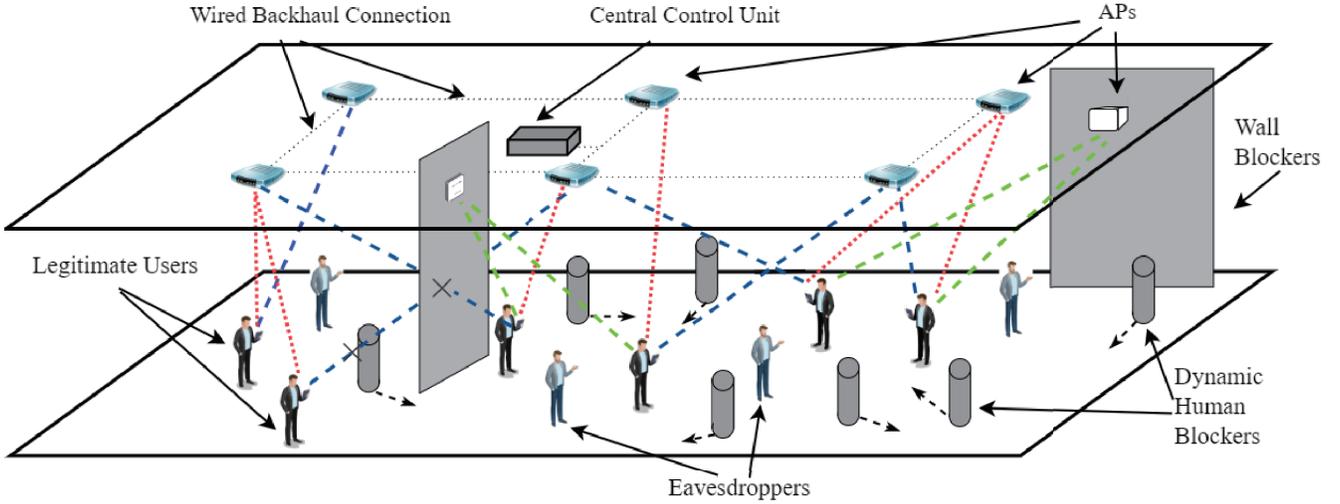}
\vspace{-2mm}
\caption{Illustration of a typical THzCom enabled indoor wireless system, where dynamic human blockers and wall blockers exist.}\label{Fig:System_Model}\vspace{-3mm}
\end{figure*}

\textbf{Sparsity:} The wavelength of THz signals are extremely short, e.g., at the level of millimeter and sub-millimeter, which is comparable to the surface roughness of typical objects in the environment. This implies that the surfaces considered smooth at lower frequencies are rough in THzCom. When THz waves hit such rough surfaces, very high reflection, diffraction, and scattering losses occur, leading to high sparsity of THz channels. The problems caused by this high sparsity, such as poor multiplexing gain, need to be properly addressed when designing effective signal processing algorithms for THzCom.

\textbf{Vulnerability to Blockages:} Given the extremely short wavelength of THz signals, THz propagation is very vulnerable to various blockages. Specifically, objects with small dimensions, e.g., a person holding a THz device, moving humans, and indoor constructions including walls and furniture, can be impenetrable blockers. This vulnerability significantly limits the distance of THz transmission and boosts the importance of line-of-sight (LoS) propagation in THzCom.

Understanding the fundamental characteristics of THz channels, including the aforementioned ones and others (e.g., temporal broadening and weather impact), is critical for integrating THzCom into 6G and beyond wireless systems. In order to design ready-to-use THzCom systems, there is an emerging need to establish generalized mathematical frameworks for analyzing key performance metrics of THzCom systems in practical and complex environment, such as the system shown in Fig. \ref{Fig:System_Model}. This need becomes more intense when we focus on massive connectivity and security, both of which have rarely been investigated.

To satisfy this need, we need to develop theoretical approaches for analyzing the user capacity of complicated THzCom systems, e.g., a system with distributed access points (APs) and a huge number of users. Here, the user capacity is defined as the maximum number of users supported by a THzCom system. It is noted that supporting simultaneous transmission towards multiple users using an LoS channel at very high frequencies is possible, if antenna spacing is much larger than the wavelength of signals. Trigger by this, the spatial degree of freedom (DoF) of THzCom systems needs to be derived as a function of dominant factors, including the number of radio frequency (RF) chains at the AP and the number of THz propagation paths in the environment \cite{Yuan_TCOM_2022}. Then, this function can be extended to incorporate other performance limiting factors, e.g., the angular spreads of channels, number and resolution of phase shifters, space between adjacent antennas, number and layout of APs, and number and properties of blockages. Once completed, we can use it jointly with existing performance evaluation frameworks, e.g., those analyzing the coverage probability of THzCom systems and the secrecy rate and secrecy outage probability of wireless systems, to reveal the trade-offs between the connectivity, security, rate, reliability, and energy and spectral efficiencies of complicated THzCom systems.

\section{Massive Connectivity in THz Communications}

\begin{figure*}[!t]
\centering
\includegraphics[width=1.8\columnwidth]{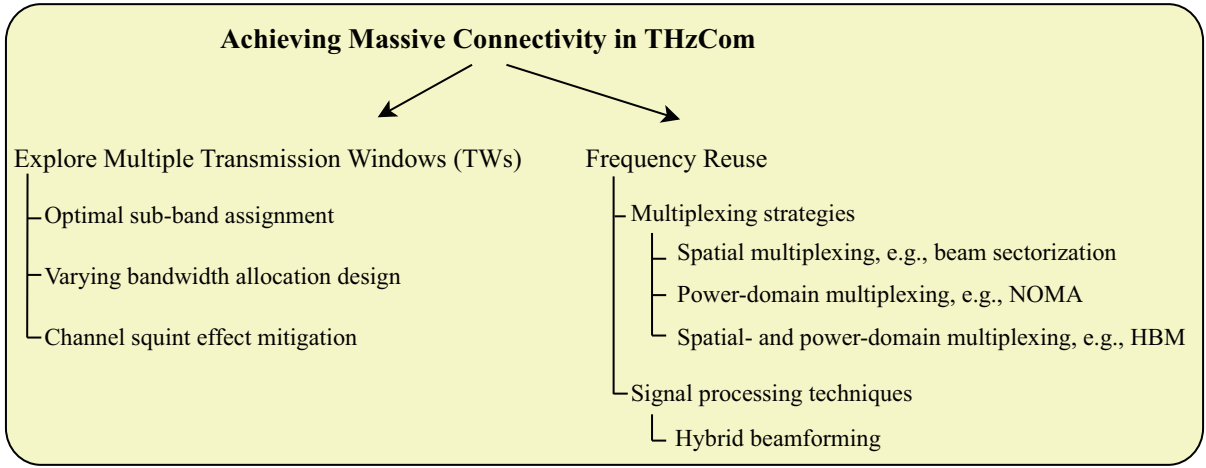}
\vspace{-2mm}
\caption{Summary of potential strategies for achieving massive connectivity in THzCom.}\label{Fig:THzCom_MC_Strategies}\vspace{-3mm}
\end{figure*}

Efficient spectrum allocation based on carried-based modulation is pivotal to harness the potential of the THz band for massive connectivity. With carrier-based modulation, the transmitted data is first modulated into single- or multi-carrier waveforms (or pulses) and then the modulated signals are upshifted to THz frequencies via frequency multipliers. Recently, the \emph{multi-band-based spectrum allocation} scheme has been discussed as a promising scheme where TWs are divided into multiple non-overlapping sub-bands. These sub-bands are used to provide ultra-high data rate transmission towards multiple users, ideally not causing interference to each other. Despite that recent studies have examined the benefit of THz spectrum allocation for massive connectivity, e.g., \cite{Yuan_TCOM_2022}, there are several pressing challenges. On one hand, the huge spectrum spanning multiple THz TWs needs to be allocated among all the users in a THzCom system. On the other hand, frequency reuse techniques need to be adopted to repeatedly use the frequency resources in a THzCom system. We next present the principles, challenges, and solutions related to using multiple TWs and frequency reuse in THzCom, where the potential massive connectivity strategies for THzCom are summarized in Fig. \ref{Fig:THzCom_MC_Strategies}.

\subsection{Exploring Multiple Transmission Windows for Massive Connectivity}

In THzCom, one fundamental challenge is caused by the high frequency selectivity of the molecular absorption loss within the spectrum spanning multiple TWs. Particularly, the variation in molecular absorption loss among the spectrum to be allocated to each user would be very high. This needs to be addressed by new solutions.

One promising solution is sub-band assignment. In multiuser THzCom systems, the distance-dependent property of the molecular absorption loss needs to be explored since users are positioned at different distances from the AP. With this exploration, distance-aware multi-carrier (DAMC) based sub-band assignment has been reported as an effective method to ensure fairness among users, since it assigns the sub-bands in the center region of the TW to the links with longer distances but the sub-bands at the edges of the TW to the links with shorter distances \cite{Han_ICC_2014}. However, it remains unanswered whether or not the DAMC based sub-band assignment is optimal when the THzCom design objective is to maximize the sum throughput and/or the number of supported users. Moreover, it is reported that the DAMC based sub-band assignment is only applicable within one TW. Thus, new problems arise when the to-be-allocated spectrum of interest spans multiple TWs. For example, the number of sub-bands in each TW needs to be optimally decided and the set of users served by each TW needs to be identified. This necessitates the optimal design of sub-band assignment. Furthermore, considering equal sub-band bandwidths, the optimal sub-band assignment can be designed as mixed-integer non-linear optimization problems and solved by adopting the transformation from binary variables to real variables and the successive convex approximation techniques. However, solving such problems can be very complex. To address this issue, non-traditional optimization techniques with low complexity need to be devised.

Another promising solution is varying bandwidth allocation. It is found that the variation in molecular absorption loss among users is very high when each user is allocated the same amount of spectrum resources (or equivalently, the same bandwidth). This variation can be reduced by allowing users to occupy varying sub-band bandwidth in spectrum allocation \cite{Akram_TCOM_2022}, as shown in Fig. \ref{Fig:THzBand}. This design introduces new problems. First, tractable expressions need to be formulated to model the molecular absorption coefficient against the frequency within the spectrum region spanning multiple TWs. Second, the spectrum allocation problem with varying sub-band bandwidth is highly non-convex, which cannot be solved using traditional optimization techniques. This necessitates the use of emerging optimization techniques such as machine learning (ML). For example, model-free unsupervised learning can be used to obtain the near-optimal solutions.

Apart from the challenges caused the molecular absorption loss, another yet-to-be-tackled challenge is to counteract the channel squint effect (CSE) at antennas. Here, the CSE refers to phase shift errors occurring in THz beamforming, since THzCom generally have a large bandwidth. To mitigate the CSE, THz true-time delays (TTDs) can be used as the alternative to phase shifters in THz digital antenna arrays, by adjusting the phase shift relative to the operating frequency.

\subsection{Frequency Reuse with Multiplexing and Signal Processing for Massive Connectivity}

Frequency reuse has great potential for THzCom systems to support a huge number of users, through adopting overlapping THz frequency resources repeatedly. With the recent advancement in THz band digital processors, it has come to light that THzCom systems can handle reasonable signal processing complexity. This opens the avenues for applying frequency reuse into THzCom, paving the way to materialize massive connectivity in the 6G and beyond era. The interference caused by frequency reuse in THzCom systems was recently analyzed \cite{Ye_TVT_2021}. Apart from performance analysis, new spatial and power-domain multiplexing strategies operated at the THz band are needed for adopting frequency reuse in THzCom systems.

One of the most encouraging spatial multiplexing strategies is beam sectorization. In a THzCom system with beam sectorization, each AP can perform beamforming to form non-overlapping beam sectors with high directional gains, and the entire THz spectrum available in the system is reused within beam sectors. In each beam sector, the spectrum can be divided into non-overlapping sub-bands which are assigned to AP-user links. Two challenges need to be tackled to reap the most benefits of beam sectorization, explained as follows:
\begin{itemize}
\item
The first challenge is inter-sector interference cancelation which minimizes or eliminates the impairment caused by non-ideal beam shapes. To tackle this challenge, we can embrace digital filter design and adopt phase antenna arrays while considering a ripple constraint on the in-sector power pattern.
\item
The second challenge is to identify sector regions, which can be addressed by two-dimensional (2D) sector deployment. Most sector deployment designs have aimed at horizontal sectorization, since vertical sectorization causes high interference if the vertical distances between the users and the AP are not comparable to the horizontal distances. Nonetheless,  THzCom systems typically have limited transmission distance; thus, the horizontal and vertical distances between the users and the AP are comparable to each other. This motivates the use of vertical sectorization in THzCom systems, while ensuring low interference. By performing horizontal and vertical sectorization, the 2D sector deployment can increase the capability of THzCom systems. With this deployment, the AP needs to deploy multiple antenna arrays with appropriate signal processing to cover $360^{\circ}$ azimuth angles and $180^{\circ}$ vertical angles. This enables each antenna array to serve a different horizontal beam sector, and within a horizontal beam sector, utilize multiple beams with different title angles. Once 2D sector deployment matured, we can blend sector deployment design with inter-sector interference cancelation in THzCom.
\end{itemize}

As an emerging power-domain multiplexing strategy, non-orthogonal multiple access (NOMA) brings an extra degree of freedom in the power domain, and its application into THzCom systems has recently attracted increasing attention, e.g., \cite{Zhang_TCOM_2021}. In NOMA, superposition coding and successive interference cancellation (SIC) are deployed to decode signals for different users in the same time and frequency resource block, offering additional access in congested traffic scenarios, e.g., ultra dense networks \cite{Ding_CM_2017}. In multiuser THzCom systems with sparse channels of users in the limited scattering environment, NOMA acts as a potential technique to improve the spectral efficiency. Specifically, NOMA can be designed to group highly correlated users into one cluster, e.g., using beam division multiple access (BDMA). Then different transmit power levels can be allocated to grouped users, allowing them to decode their signals using advanced SIC. A key challenge in NOMA-aided THzCom systems is to construct effective clustering criteria, including full channel knowledge based criterion and location based criterion, and evaluate their impacts on the system spectral efficiency.

Given the benefits of spatial multiplexing and power-domain multiplexing, a natural question arises: \emph{Is there a paradigm to realize the joint benefits of these two multiplexing strategies?} Our answer is ``Yes, the hierarchical bandwidth modulation (HBM)''. Exploring the distance varying usable bandwidth property of THz TWs, HBM enables simultaneously transmitting multiple data streams over links with different distances and within the same THz TW, by adapting both symbol duration and modulation order \cite{Hossain_ICC_2019}. Specifically, the symbol duration of the links with shorter distances is shorter than that of the links with longer distances.

Different from NOMA where the same spectrum bandwidth is used by different users, users in HBM adopt their respective usable bandwidth within TWs, and the usable bandwidth varies among users as long as their transmission distances change. Thus, in order for HBM to be effective, there should be reasonable differences in the usable bandwidth among different users. This is indeed possible when the transmission distances of users are different. Some open problems related to HBM are: 1) How can we optimally select symbol duration and modulation order when the number of transmission links is large and the usable bandwidths of different links are similar? 2) How can we analyze the impact of temporal broadening on HBM, since signals modulated in HBM exhibit high frequency selectivity? 3) How can we identify the optimal usable bandwidth within the TW for each user? The third question is critical since in the literature, the spectrum region with less than $3~\textrm{dB}$ path loss difference is defined as the usable bandwidth. However, the optimality of this definition is not guaranteed. Hence, the usable bandwidth within the TW for each user needs to be optimally determined in HBM-aided THzCom systems.

\begin{figure}[t]
\centering
\includegraphics[width=0.9\columnwidth]{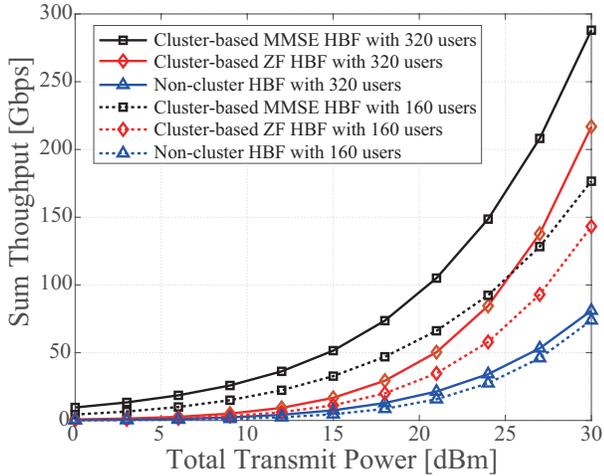}
\vspace{0mm}
\caption{Sum throughput of three HBF schemes,  namely, cluster-based minimum mean squared error (MMSE) HBF, cluster-based zero-forcing (ZF) HBF, and non-cluster HBF, in a THzCom system where a 256-antenna AP with 16 RF chains serves 160 or 320 users.}\label{Fig:Connectivity}
\vspace{-3mm}
\end{figure}

Together with multiplexing, advanced signal processing techniques can further improve the connectivity performance of THzCom systems. Focusing on the transmission from APs to users, new multi-carrier hybrid beamforming (HBF) need to be developed at APs, no matter they are co-located or distributed, to maximize the number of served users. This development requires us to devise effective techniques including beamwidth control, dynamic antenna selection, and adaptive power allocation. Furthermore, ML techniques, such as deep neural network based optimization, can be used to determine the optimal beamforming parameters in complex environment, ensuring the maximum user capacity being achieved. A preliminary result of using HBF with multiplexing to serve a very large number of users in THzCom is shown in Fig. \ref{Fig:Connectivity}, where users are grouped into different clusters. Multiple users in one cluster is served by a single RF chain and different clusters are divided by BDMA with various HBF schemes. With such operations, a large number of users can be supported by a limited number of RF chains in THzCom.

\section{Secrecy in THz Communications}

Security is an exciting and challenging problem in THzCom systems. Specifically, the information sent by a legitimate transmitter is received by not only legitimate receivers but unintended receivers (referred to as eavesdroppers). Thus, establishing effective secure communications is a testament to the success of THzCom systems. Over the past decade, physical layer security has been recognized as a promising and low-complexity paradigm to safeguard wireless systems by nicely complementing cryptographic techniques \cite{Yang_CM_2015}. As per the principles of physical layer security, wireless systems need to simultaneously improve the quality of signal received at legitimate users and diminish the quality of signal received at eavesdroppers. Although the strong directivity of THz beam brings inherent benefits to wireless security, the physical layer security design in THzCom is still challenging, especially in the scenario where eavesdroppers exist within the transmitter's beam sector or near legitimate users. We next present spectrum allocation and signal processing strategies to enhance the security of THzCom systems.

\subsection{Spectrum Allocation for Security}

\begin{figure}[t]
\centering
\includegraphics[width=0.98\columnwidth]{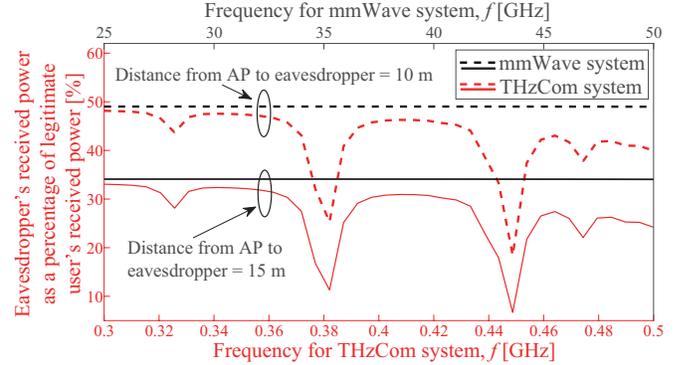}
\vspace{0mm}
\caption{Frequency- and distance-dependent eavesdropper's received power as a percentage of legitimate user's received power when the distance from AP to legitimate user is 5m.}\label{Fig:Security}
\vspace{-3mm}
\end{figure}

Secure THzCom can be achieved by exploring the abundant spectrum availability at the THz band, as well as the frequency- and distance-dependent nature of molecular absorption loss within the abundant spectrum. Specifically, we can divide the abundant spectrum at the THz band into a huge number of sub-bands and then dynamically select a certain number of sub-band for secure transmission, based on the distance between the AP and legitimate users and the distance between the AP and eavesdroppers. This forms the distance-adaptive absorption peak modulation (DA-APM) scheme \cite{Gao_TWC_2020}. Given that the molecular absorption loss depends on frequency and distance, the path loss variation with distance in one sub-band is likely to be different from that in another sub-band. Hence, a well-thought-out selection of sub-bands can achieve a small path loss at the legitimate user but a large path loss at the eavesdropper, which enlarges the difference in signal quality between them and in turn improves the secrecy, as shown in Fig. \ref{Fig:Security}. Specifically, the eavesdropper's received signal power is less than 50\% of the legitimate user's received signal power, and can be less than 10\% at some frequencies. This shows the potential of spectrum allocation for secrecy enhancement. Notably, it is very difficult for the eavesdropper to intervene sub-band selection, as the selection is made based on distances, unless it moves closer to the legitimate user which is easy to identify.

Despite the potential of the DA-APM scheme has been touched, there are still open challenges to safeguard THzCom systems. By looking closely at the DA-APM scheme, we find that the sub-band selection problem, especially for the THzCom system with a large number of users, can be a mixed-integer non-linear problem. To solve this problem, we can resort to binary-to-integer variable transformation, together with successive convex approximation, for attaining near-optimal solutions. Also, the DA-APM scheme only considers equal bandwidth of sub-bands. If this consideration can be relaxed by varying the bandwidth of sub-bands, a more dramatic change in path loss will occur within sub-bands, which brings higher security when the eavesdropper is positioned farther than the legitimate user. Furthermore, to effectively implement the DA-APM scheme, new THz signal design with smooth power spectral density is required to avoid signal leakage to the eavesdropper.

It is noted that the DA-APM scheme is only valid when the AP-eavesdropper distance is longer than the AP-user distance. When the eavesdropper exists between the AP and the legitimate user, additional signal processing methods were proposed to improve the secrecy performance of THzCom systems, such as adding artificial noise (AN) at the full-duplex legitimate user \cite{Gao_TWC_2022}. When doing so, the AN received at the eavesdropper causes interference to its received signal from the AP, due to the temporal broadening effect (TBE). Here, the TBE means the widening of transmitted pulses caused by the frequency-dependent nature of molecular absorption loss. Particularly, the TBE increases exponentially with the transmission distance and the molecular absorption coefficient within each sub-band. Trigger by this property, we propose to adopt varying pulse widths within each sub-band to further enhance the secrecy. By doing so, the time interval between two consecutive pulses within randomly selected sub-bands can be carefully set, to ensure non-overlapping broadened pulses being detected at the legitimate user but overlapping broadened pulses being detected at the eavesdropper, which enhances the secrecy performance. In addition, the existing secure transmission schemes do not consider practical hardware imperfections such as transmitter and filter non-linearities, I/Q imbalance, and phase shift errors. This requires practice oriented design and experimental validation of secure THzCom schemes.

\subsection{Signal Processing for Security}

It is worthwhile mentioning that the existing secure communication schemes discussed in the literature cannot be applied when eavesdroppers exist at the close proximity of legitimate users. To overcome this bottleneck, frequency diverse array (FDA) based beamforming has emerged as an attractive solution. At the FDA, linear phase progression is adopted over the aperture and induces an electronic beam scan. When applying an additional linear frequency shift, the FDA generates a new term which brings a scan angle that varies with the range of the far-field. This intentionally introduces frequency offsets among array antennas and decouples the highly correlated channels of legitimate users and eavesdroppers. Such decoupling capability allows the FDA to degrade the received signal quality at eavesdroppers and accordingly enhance the secrecy performance.

There are several open questions to reap the potential of FDA based beamforming. First, there is a clear need to develop the hardware of efficient FDAs. In the THz band, widely spaced arrays can act as FDAs due to their range-dependent array pattern. Thus, an effective design of widely spaced arrays can group antennas into several subarrays and separate these subarrays over hundreds of wavelengths to reduce their correlation. Second, the comparison between linear FDAs and random FDAs needs to be conducted. Using linear FDAs, the achieved direction and range are coupled. Thus, multiple direction-range pairs may exist such that the eavesdropper receives the same signals as the legitimate user, impairing secure transmission. To tackle this problem, random FDAs can be utilized where each transmit antenna is randomly, but not linearly, allocated a narrow- or sub-band frequency. Thus, random FDAs decouple the correlation between direction and range, making itself promising for robust secure THzCom \cite{Hu_Access_2017}. Third, the far-field beam pattern produced by FDAs is time variant, as the frequency increments between adjacent antennas are small. This time-variant property may enable short-time secure transmission, which can be used to safeguard the transmission of control signals between an AP and legitimate users, e.g., in a smart factory. To support long-time secure transmission of data signals, this time-variant property needs to be addressed. Fourth, advanced signal processing algorithms, such as transmitter AN and confidential broadcasting, can be jointly used with FDAs to further improve the secrecy performance, where its memory requirement may be examined.

\section{Conclusion}

This article has provided an overview and comprehensive discussions on massive connectivity and security issues in THzCom systems by fully exploiting the challenges and opportunity brought by THz spectrum. First, the establishment of new performance analytical frameworks under the key characteristics of THz signal propagation has been revealed. Then, new and effective approaches, in terms of spectrum management and signal processing, have been discussed to increase the number of supported users and enhance the secrecy performance of THzCom systems. In such discussion, critical challenges and promising research directions have been presented, offering insightful guidance to develop THzCom systems accommodating massive connectivity and promoting security.

\section*{Biographies}

\noindent \footnotesize \textbf{Nan Yang} is an Associate Professor with the School of Engineering at the Australian National University, Australia. His research interests include THz communications, ultra-reliable low latency communications, cyber-physical security, and molecular communications.

\vspace{2mm}

\noindent \textbf{Akram Shafie} is currently pursuing the Ph.D. degree with the School of Engineering at the Australian National University, Australia. His research interests include THz and mmWave communications and machine learning in communications.

\end{document}